\documentclass[twocolumn,prb,showpacs,superscriptaddress,floatfix,nofootinbib]{revtex4}
\usepackage{amssymb}
\usepackage{amsmath}
\usepackage{graphicx}

 \DeclareMathOperator{\sgn}{\mathrm{sgn}}
 \DeclareMathOperator{\Real}{\mathrm{Re}}

\begin{document}

\title{Density of states anomalies in hybrid
superconductor-ferromagnet-normal metal structures}
\author{A.~A.~Golubov}
\affiliation{Faculty of Science and
Technology, University of Twente, The Netherlands}
\author{M.~Yu.~Kupriyanov}
\affiliation{Nuclear Physics Institute, Moscow State University, 119992 Moscow, Russia}
\author{M.~Siegel}
\affiliation{Institute for Micro and Nanoelectronic Systems, \ \ Karlsruhe University,
Hertzstrasse 16, D-76187 Karlsruhe, Germany}
\date{20 January 2005}

\begin{abstract}
The results of calculations of the spatially-resolved density of
states (DoS) in an S(F/N) bilayer are presented (S is a
superconductor, F is a metallic ferromagnet, N is a normal metal)
within quasiclassical theory in the dirty limit. Analytical
solutions are obtained in the case of thin F, N layers which
demonstrate the peculiar features of DoS in this system. The
dependencies of the minigap and the DoS peak positions on the
exchange energy and parameters of the layers are studied
numerically.
\end{abstract}

\maketitle

In the past few years there was a noticeable interest to the Josephson
junctions with ferromagnetic barriers due to possibility to realize the $\pi
-$junctions having the phase difference $\pi $ in the ground state. The $\pi
-$states in SFS junctions were first predicted by \cite%
{Buzdin,Buzdin1,Buzdin2} and realized experimentally by Ryazanov \textit{et
al}. \cite{Ryazanov,Ryazanov1} in Nb/CuNi/Nb structures and later by other
groups \cite{Kontos,Blum,Bourgeois,Surgers,Sellier} using different
ferromagnetic barriers. These experiments stimulated further theoretical
activity (see \cite{GKI} for the review). In particular, Josephson
structures composed from arrays of $0-$ and $\pi -$ Josephson junctions
should exhibit extraordinary characteristics \cite{Mints,BuzKosh}. Such
arrays were recently realized in zig-zag HTS/LTS structures \cite{Hilgen}.

The purpose of the present paper is to study spatially resolved electronic
density of states (DoS) in the structure of S(FN)type, consisting of a bulk
superconductor with ferromagnetic and normal layers on the top of it, which
is a generic system for 0- and $\pi -$junctions connected in parallel.

DoS in SF bilayers (a ferromagnet coupled to a superconductor) was studied
quite extensively before. Two new features were predicted compared to SN
systems: spin splitting and spatial oscillations of DoS in a ferromagnet
\cite{Buzdin3,Buzdin4,Zareyan,Bergeret,Kriv,Golubov1,Huertas}. The effect of
spatial oscillations was quite extensively discussed in the theoretical
literature in different models \cite{Buzdin3,Buzdin4,Zareyan,Bergeret} and
observed experimentally \cite{Kontos1}. This effect is closely related to 0-$%
\pi $ transitions. The effect of splitting is relevant for thin
ferromagnetic layers and was studied theoretically in \cite{Kriv,Golubov1}.
In the present work we discuss an interplay between the oscillations and
splitting in a more complex S(F/N) structure.

\begin{figure}[tbp]
\centerline{\includegraphics[width=85mm]{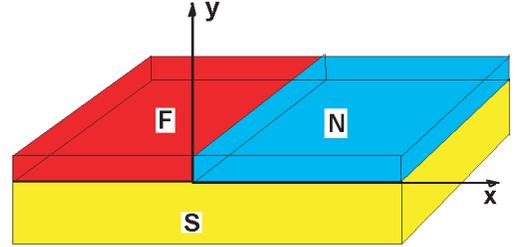}}
\caption{The geometry of the structure.}
\end{figure}

The geometry of the structure is shown on Fig.1. We assume that the dirty
limit conditions are fulfilled in all metals, F is a weak monodomain
ferromagnet with the exchange energy $H$ much smaller than the Fermi energy
and the interfaces are not magnetically active. In this case spin dependent
corrections to the resistivities can be neglected and the S(F/N) structure
is described by the following spin independent suppression parameters%
\begin{equation}
\gamma _{BF}=R_{BF}\mathcal{A}_{BF}/\rho _{F}\xi _{F},\quad \gamma _{F}=\rho
_{S}\xi _{S}/\rho _{F}\xi _{F},  \label{gammasF}
\end{equation}%
\begin{equation}
\gamma _{BN}=R_{BN}\mathcal{A}_{BN}/\rho _{N}\xi _{N},\quad \gamma _{N}=\rho
_{S}\xi _{S}/\rho _{N}\xi _{N},  \label{gammasN}
\end{equation}%
\begin{equation}
\gamma _{B}=R_{B}\mathcal{A}_{B}/\rho _{N}\xi _{N},\quad \gamma =\rho
_{F}\xi _{F}/\rho _{N}\xi _{N}.  \label{gammasFN}
\end{equation}%
Here $R_{BF},$ $R_{BN},$ $R_{B}$ are the specific resistivities of the SF,
SN and NF interfaces respectively; $\rho _{S,F,N}$ , $D_{S,F,N}$ and $\xi
_{S,F,N}$ are the resistivities, the diffusion constants and the coherence
lengths of the S, F and N layer and the coherence lengths, where $\xi
_{S,F,N}=\sqrt{D_{S,F,N}/2\pi T_{c}}$ and $T_{c}$ is the critical
temperature of the superconductor.

Under the above assumptions the problem can be solved in the framework of
the Usadel equations \cite{Usadel}. To simplify it further we assume that S
is a bulk superconductor and $\gamma _{N}\ll \gamma _{BN},\quad \gamma
_{F}\ll \gamma _{BF}$ so that the rigid boundary conditions
\begin{equation*}
F_{S}=\frac{\Delta }{\sqrt{\omega ^{2}+\Delta ^{2}}},\quad G_{S}=\frac{%
\omega }{\sqrt{\omega ^{2}+\Delta ^{2}}}
\end{equation*}%
are valid for superconductor. Here $\Delta $ is the magnitude of the order
parameter in S electrode, $F_{S}$ and $G_{S}$ are the Green's functions, $%
\omega =\pi T(2n+1)$ are the Matsubara frequencies.

Let us choose the $x,y$ axes as shown in Fig.1 and use the $\theta $
parametrization $G=\cos \theta ,$ $F=\sin \theta $, then the Usadel
equations have the form%
\begin{equation}
\xi _{F,N}^{2}\frac{\pi T_{c}}{\widetilde{\omega }}\left\{ \frac{\partial
^{2}}{\partial x^{2}}\theta _{F,N}+\frac{\partial ^{2}}{\partial y^{2}}%
\theta _{F,N}\right\} -\sin \theta _{F,N}=0,  \label{EqUSF}
\end{equation}%
where $\widetilde{\omega }=\omega +iH$ in F and $\widetilde{\omega }=\omega $
in N.

The boundary conditions at the SF $(y=0,$ $-\infty <x\leq 0)$, SN $(y=0,$ $%
0\leq x<\infty )$ and FN $(x=0,$ $0\leq y\leq d_{F},$ $d_{N})$ interfaces
have the form \cite{KL}

\begin{equation}
\gamma _{B(F,N)}\xi _{F,N}\frac{\partial }{\partial y}\theta _{F,N}=-\sin
(\theta _{S}-\theta _{F,N}),~y=0,  \label{BC_Fi_F}
\end{equation}

\begin{gather}
\gamma _{B}\xi _{F}\frac{\partial }{\partial x}\theta _{F}=\sin (\theta
_{N}-\theta _{F}),~x=0,~0\leq y\leq d_{F},d_{N},  \label{BC_Fi_NF} \\
\xi _{N}\frac{\partial }{\partial x}\theta _{N}=\gamma \xi _{F}\frac{%
\partial }{\partial x}\theta _{F},~x=0,~0\leq y\leq d_{F},d_{N},  \notag
\end{gather}%
where $\sin \theta _{S}=\Delta /\sqrt{\omega ^{2}+\Delta ^{2}}$ and $\cos
\theta _{S}=\omega /\sqrt{\omega ^{2}+\Delta ^{2}}$. At the free interfaces
the boundary conditions are%
\begin{equation}
\frac{\partial }{\partial y}\theta _{F,N}=0,\quad y=d_{F,N}  \label{BC_free1}
\end{equation}%
\begin{equation}
\frac{\partial }{\partial x}\theta _{F,N}=0,\quad 0\leq y\leq d_{F},\quad
x\rightarrow \mp \infty  \label{BC_min}
\end{equation}

We will consider the limit of thin F and N layers $d_{F,N}\ll \xi _{F,N}$.
In this case one can neglect both the derivative on $x$ and nongradient
items in Usadel equations (\ref{EqUSF}) and substitute the resulting
solutions%
\begin{equation*}
\theta _{F,N}(x,y)=\theta _{F,N}(x)-K_{F,N}(x)\frac{d_{F,N}}{\xi _{F,N}^{2}}%
y+K_{F,N}(x)\frac{y^{2}}{2\xi _{F,N}^{2}}
\end{equation*}%
\begin{equation*}
K_{F,N}(x)=\left\{ \frac{\widetilde{\omega }}{\pi T_{c}}\sin \theta
_{F,N}-\xi _{F,N}^{2}\frac{\partial ^{2}}{\partial x^{2}}\theta
_{F,N}\right\} \frac{d_{F,N}}{\xi _{F,N}^{2}}
\end{equation*}%
into boundary conditions (\ref{BC_Fi_F}). Then the problem is reduced to the
one-dimensional equations for lateral variations of $\theta _{N,F}$\ in the $%
x$-direction:

\begin{equation}
\zeta _{N,F}^{2}\frac{\partial ^{2}}{\partial x^{2}}\theta _{N,F}-\sin
(\theta _{N,F}-\theta _{N,F}(\pm \infty ))=0,  \label{TetNx}
\end{equation}%
where
\begin{equation}
\theta _{N,F}(\pm \infty )=\arctan \frac{\pi T_{c}\sin \theta _{S}}{%
\widetilde{\omega }\tilde{\gamma}+\pi T_{c}\cos \theta _{S}},
\label{TetFinf}
\end{equation}%
the decay lengths $\zeta _{N}$\ and $\zeta _{F}$\ are%
\begin{equation}
\zeta _{N,F}=\xi _{N,F}\sqrt{\frac{\pi T_{c}\tilde{\gamma}\cos \theta
_{N,F}(\pm \infty )}{(\widetilde{\omega }\tilde{\gamma}+\pi T_{c}\cos \theta
_{S})}},  \label{ksiN}
\end{equation}%
and we have taken for simplicity equal barrier parameters for F and N

\begin{equation}
\gamma _{BN}\frac{d_{N}}{\xi _{N}}=\gamma _{BF}\frac{d_{F}}{\xi _{F}}\equiv
\widetilde{\gamma }.  \label{gammas}
\end{equation}

The general solution of Eq. (\ref{TetNx}) has the form%
\begin{equation}
\theta _{N,F}(x)=\theta _{N,F}(\pm \infty )+  \label{SolTet_N}
\end{equation}%
\begin{equation*}
+4\arctan \left[ (\tan \frac{\theta _{N,F}(0)-\theta _{N,F}(\pm \infty )}{4}%
)\exp \{\mp \frac{x}{\zeta _{N,F}}\}\right] ,
\end{equation*}%
The integration constants $\theta _{N}(0)$ and $\theta _{F}(0)$ in ( \ref%
{SolTet_N}) have to be determined from the boundary conditions (\ref%
{BC_Fi_NF}) at $x=0$ and can be found analytically in the limit of large
transparency of the FN interface when $\theta (x)$ is continuous \emph{at at}
$x=0$

\begin{equation*}
\theta _{N}(0)=\theta _{F}(0)=\theta (0).
\end{equation*}

From (\ref{ksiN}) it follows that the effective decay length in normal
metal, $\zeta _{N},$ is a real quantity and equals to $\zeta _{N}=\xi _{N}%
\sqrt{\widetilde{\gamma }}$ for small $\omega $ and tends to $\zeta _{N}=\xi
_{N}\sqrt{\pi T_{c}/\omega }$ with $\omega $ increase. The effective decay
length $\zeta _{F}$ in ferromagnet and at low $\omega \ll \Delta ,H/%
\widetilde{\gamma }$ is given by $\zeta _{F}=\xi _{F}\sqrt{\widetilde{\gamma
}/\sqrt{1-\widetilde{\gamma }^{2}(H/\pi T_{c})^{2}}}$\emph{\ }i. e. it
becomes complex for sufficiently strong exchange field $H>\pi T_{c}/%
\widetilde{\gamma }.$

Below we consider several limiting cases.

\textbf{Identical F and N metals. }Assume for simplicity that the F and N
materials differ by the existence the exchange field in F ($\gamma =1,$\ $%
\xi _{F}=\xi _{N}=\xi $), then from (\ref{BC_Fi_NF}) for $\theta (0)$ we
have
\begin{equation}
\theta (0)=2\arctan \frac{\sin \frac{\theta _{N}(\infty )}{2}+g\sin \frac{%
\theta _{F}(-\infty )}{2}}{\cos \frac{\theta _{N}(\infty )}{2}+g\cos \frac{%
\theta _{F}(-\infty )}{2}},\quad g=\frac{\zeta _{N}}{\zeta _{F}}
\label{Tet(0)1}
\end{equation}

\textbf{Identical F metals with antiparallel direction of magnetization}%
\textit{. }The results can be easily generalized to the case of an S(F/F)
structure with two identical ferromagnetic films having opposite
magnetization directions (antiferromagnetic configuration)
\begin{equation}
\theta (0)=2\arctan \frac{g^{\ast }\sin \frac{\theta _{F}^{\ast }(-\infty )}{%
2}+g\sin \frac{\theta _{F}(-\infty )}{2}}{g^{\ast }\cos \frac{\theta
_{F}^{\ast }(-\infty )}{2}+g\cos \frac{\theta _{F}(-\infty )}{2}},
\label{Tet(0)Anti}
\end{equation}

Using the solutions obtained above one can calculate the spatially resolved
DoS in S(F/N) and S(F/F) structures.

\textbf{DoS in S(F/N) and S(F/F) proximity systems. }The DoS for each spin
direction is given by
\begin{equation}
N(\varepsilon )=\frac{N_{0}}{2}\Real G(\omega \rightarrow -i\varepsilon
+\delta ),  \label{DOS_G}
\end{equation}%
where $N_{0}$ is the total DoS for both spins at the Fermi surface in the
normal state and $G(\varepsilon -i\delta )=\cos \theta (\varepsilon -i\delta
)$ is the retarded Green's function. The total DoS is found by summing over
both spin projections i.e. $N^{total}=N(H)+N(-H)$.

\textbf{DoS in N and F metals far from the F/N interface.} In a normal metal
far from the F/N interface ($x=\infty $) the total DoS is given by%
\begin{gather}
N_{N}(\varepsilon )=N_{0}\Real\frac{\widetilde{\varepsilon }_{N}}{\Omega _{N}%
},\quad \Omega _{N}=\sqrt{\widetilde{\varepsilon }_{N}^{2}-\Delta ^{2}}\sgn%
(\varepsilon )  \label{DENS_N} \\
\widetilde{\varepsilon }_{N}=\varepsilon (1+\widetilde{\gamma }\sqrt{\Delta
^{2}-\varepsilon ^{2}}/\pi T_{c}).  \notag
\end{gather}%
It is well known (see Refs.\cite{MCM,GK89}) that DoS in a F/N bilayer has a
minigap at $\varepsilon _{g}<\Delta $, which depends on the value of $%
\widetilde{\gamma }$, and $N_{N}(\varepsilon )$ has the peaks at $%
\varepsilon =\varepsilon _{g}$ and $\varepsilon =\Delta $. The minigap $%
\varepsilon _{g}$ characterizes the strength of superconducting correlations
induced into N metal due to the proximity effect.

\bigskip
\begin{figure}[tbp]
\centerline{\includegraphics[width=85mm]{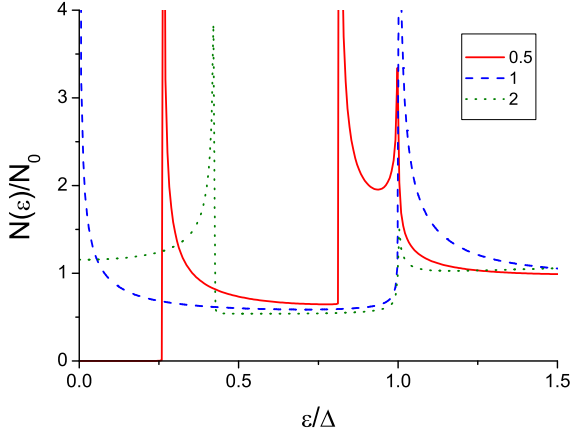}}
\caption{The total DoS in an SF bilayer for various values of $\widetilde{%
\protect\gamma }h$ as indicated in the figure.}
\end{figure}

In SF bilayers, modifications of DoS due to spin splitting of energy levels
were investigated in Refs.\cite{Kriv,Golubov1}. The DoS per spin projection
in the F layer has the form%
\begin{gather}
N_{F\uparrow ,F\downarrow }(\varepsilon )=\frac{N_{0}}{2}\Real\frac{%
\widetilde{\varepsilon }_{F\uparrow ,F\downarrow }}{\Omega _{F\uparrow
,F\downarrow }},  \label{DENS} \\
\Omega _{F\uparrow ,F\downarrow }=\sqrt{\widetilde{\varepsilon
}_{F\uparrow
,F\downarrow }^{2}-\Delta ^{2}}\sgn(\varepsilon \mp H), \notag \\
\widetilde{\varepsilon }_{F\uparrow ,F\downarrow }=\varepsilon +\widetilde{%
\gamma }(\varepsilon \mp H)\sqrt{\Delta ^{2}-\varepsilon ^{2}},
\notag
\end{gather}%
which demonstrates the energy renormalization due to the exchange
field. In particular, it follows from (\ref{DENS}) that now there
are two minigaps in the spectrum\ $\varepsilon _{g\uparrow }$\ and\
$\varepsilon _{g\downarrow }$ \ and\ $\varepsilon _{g\uparrow }\leq
\varepsilon _{g}\leq \varepsilon _{g\downarrow }$.

The total DoS in a F/N bilayer $N_{tot}(\varepsilon )=$$N_{F\downarrow
}(\varepsilon )+N_{F\uparrow }(\varepsilon )$ is shown in Fig.2. It is
clearly seen that at $h=H/\pi T_{c}<1/\widetilde{\gamma }$ there are three
peaks in DoS located at $\varepsilon _{g\downarrow },$\ $\varepsilon
_{g\uparrow }$and\ $\Delta $ respectively. At $h=1/\widetilde{\gamma }$\ the
low energy singularity is shifted to the Fermi level and for $h>1/\widetilde{%
\gamma }$ the first peak disappears resulting in only two singularities in
the DoS at $\varepsilon =\varepsilon _{g\uparrow }$and\ $\varepsilon =\Delta
.$Note that the total DoS at low energies depends nonmonotonously on $H$\
even in a thin F-layer, even though spatial oscillations are absent across
the layer. Eq. (\ref{DENS}) yields $N_{_{F\uparrow ,F\downarrow
}}(\varepsilon =0)=(N_{0}/2)\Real\widetilde{\gamma }h\sgn(h)/\sqrt{%
\widetilde{\gamma }^{2}h^{2}-1})$. For $\widetilde{\gamma }h<1$\ the total
DoS $N(0)=0$\ due to the minigap in F while for $\widetilde{\gamma }h\geq 1$%
\ the total low-energy DoS increases sharply, exceeds unity and saturates at
$N(0)=N_{0}$\ for l $\widetilde{\gamma }h\gg 1.$

\textbf{DoS at the F/N interface. }At $x=0$and for identical transport
parameters on the F and N metals from (\ref{Tet(0)1}), (\ref{DOS_G}) we
obtain%
\begin{equation}
N_{F\uparrow ,F\downarrow }(\varepsilon )=\frac{N_{0}}{2}\Real\frac{-i%
\widetilde{\varepsilon }_{N}-i\widetilde{\varepsilon }_{F\uparrow
,F\downarrow }+2\widetilde{\varepsilon }_{FN}}{\Omega _{F\uparrow
,F\downarrow }+\Omega _{N}+2\sqrt{\Delta ^{2}-\widetilde{\varepsilon }%
_{FN}^{2}}},  \label{DENS0}
\end{equation}%
where%
\begin{equation}
\widetilde{\varepsilon }_{FN}=\sqrt{\frac{\Omega _{N}\Omega _{F\uparrow
,F\downarrow }-\widetilde{\varepsilon }_{N}\widetilde{\varepsilon }%
_{F\uparrow ,F\downarrow }-\Delta ^{2}}{2}},  \label{efn}
\end{equation}%
and $\Omega _{N},\widetilde{\varepsilon }_{N}$\ and $\Omega _{F\uparrow
,F\downarrow },\widetilde{\varepsilon }_{F\uparrow ,F\downarrow }$\ are
defined by (\ref{DENS_N}) and (\ref{DENS}) respectively.

\begin{figure}[tbp]
\centerline{\includegraphics[width=85mm]{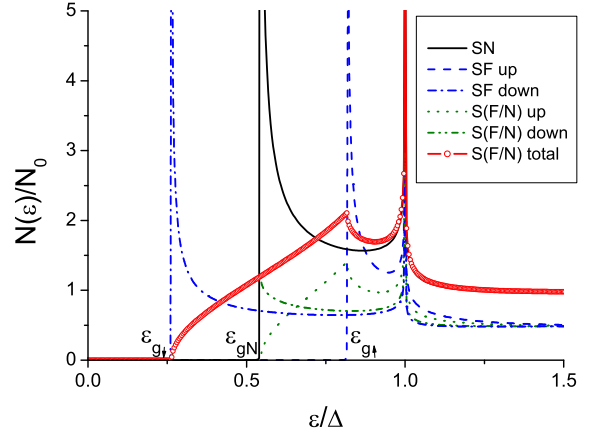}}
\caption{Spin resolved DoS: comparison of FS, FN and S(FN) for $h\widetilde{%
\protect\gamma }$=0.5.}
\end{figure}

It follows from Eq.(\ref{DENS0}) that, similar to the case of the SF bilayer
considered above, the minigap exists if $\widetilde{\gamma }h<1$. With
increasing exchange field the total DoS at $\varepsilon =0$ becomes nonzero
if $\widetilde{\gamma }h>1$ at is given by simple expression
\begin{equation}
N(0)=N_{0}\sqrt{\widetilde{\gamma }^{2}h^{2}-1}/\widetilde{\gamma }h.
\label{ZG}
\end{equation}

\bigskip \textbf{DoS at the F/F interface. }At $x=0$ \ we have%
\begin{equation}
N(\varepsilon )/N_{0}=\Real\frac{-2i\widetilde{\varepsilon }_{N}+2\widetilde{%
\varepsilon }_{FF}}{\Omega _{F\uparrow }+\Omega _{F\downarrow }+2\sqrt{%
\Delta ^{2}-\widetilde{\varepsilon }_{FF}^{2}}}  \label{den_FF}
\end{equation}%
\begin{equation*}
\widetilde{\varepsilon }_{FF}=\sqrt{\frac{\Omega _{F\uparrow }\Omega
_{F\downarrow }-\widetilde{\varepsilon }_{F\uparrow }\widetilde{\varepsilon }%
_{F\downarrow }-\Delta ^{2}}{2}}
\end{equation*}

It can be shown that DoS at the F/F interface given by Eq.(\ref{den_FF})
coincides exactly with the total DoS for the F/N interface, $N_{F\uparrow
}(\varepsilon )+N_{F\downarrow }(\varepsilon )$, where $N_{F\uparrow
}(\varepsilon )$ and $N_{F\downarrow }(\varepsilon )$ are given by Eq.(\ref%
{DENS0}). In particular, the minigap exists if $\widetilde{\gamma }h\leq 1$
and at $\widetilde{\gamma }h>1$ the DoS at F/F is determined by Eq.(\ref{ZG}%
).

\begin{figure}[tbp]
\centerline{\includegraphics[width=85mm]{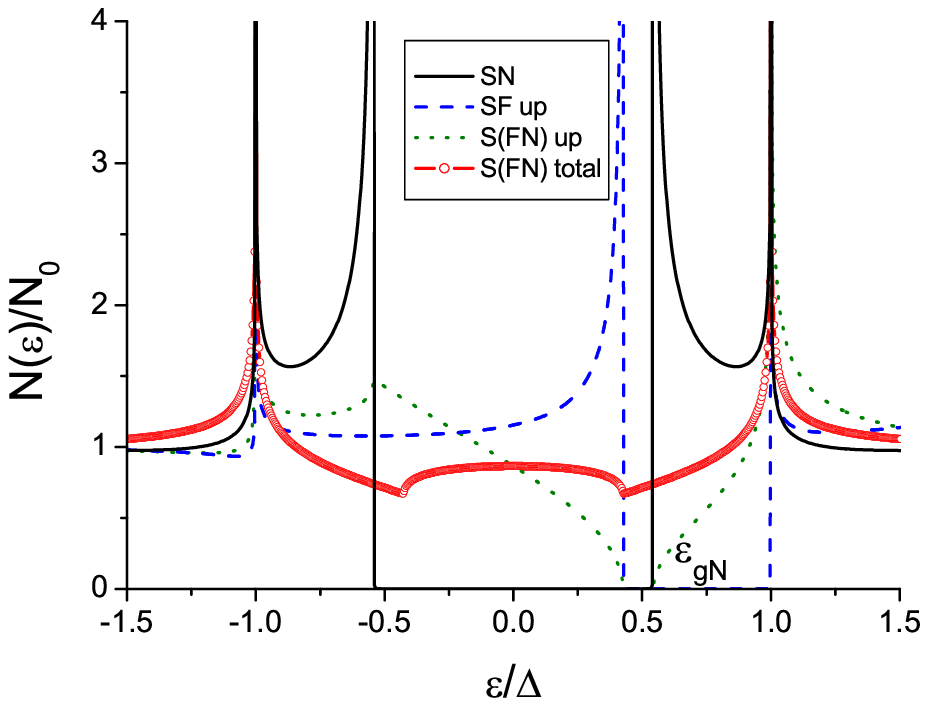}}
\caption{Spin resolved DoS: comparison of FS, FN and S(FN) for $h\widetilde{%
\protect\gamma }$=2.}
\end{figure}

The results of calculations from Eq.(\ref{DENS0}) at low temperatures $T\ll
T_{c}$ are shown in Figs.3,4 for $h\widetilde{\gamma }=0.5$ and $h
\widetilde{\gamma }=2$, respectively, together with the DoS for SF $%
(x\rightarrow -\infty )$\ and SN $(x\rightarrow \infty )$\ bilayers.

There are four characteristic energies in the system: $\varepsilon
_{g\downarrow },$ $\varepsilon _{g},$ $\varepsilon _{g\uparrow }$ and $%
\Delta .$ Here $\varepsilon _{g\downarrow }$ is the minigap for the
spin-down subband SF bilayer at $x\rightarrow -\infty $. It follows from Eq.(%
\ref{DENS0}) that $N_{F\downarrow }(\varepsilon )=0$ at $\varepsilon \leq
\varepsilon _{g\downarrow }$ and becomes nonzero at $\varepsilon
>\varepsilon _{g\downarrow }$, i.e. $\varepsilon _{g\downarrow }$ is the
minigap for the spin-down subband in S(FN) at $x=0.$ However, contrary to SF
case $N_{F\downarrow }(\varepsilon )$ has no peak $\varepsilon =\varepsilon
_{g\downarrow }$ but grows continuously from zero value.

For the spin-up subband, the minigap in $N_{F\uparrow }(\varepsilon )$ is
not equal to the gap $\varepsilon _{g\uparrow }$ in the spin-up subband in
SF bilayer at $x\rightarrow -\infty $. Instead, $N_{F\uparrow }(\varepsilon
) $ the gap value is determined by $\varepsilon _{g}$, the minigap in SN
bilayer at $x\rightarrow \infty .$ The formal reason is that in the interval
$\varepsilon \geq \varepsilon _{g} $, $\Omega _{N}$ becomes an imaginary
number and both numerator and denominator in Eq.(\ref{DENS0}) are complex
thus leading to nonzero DoS in this energy range. Similar to the spin-down
case, there is no peak in $N_{F\uparrow }(\varepsilon )$ at the gap energy $%
\varepsilon =\varepsilon _{g}$, while the peak occurs at $\varepsilon
=\varepsilon _{g\uparrow }$ (see Fig.3). With further increase of energy
there is a sharp peak in DoS at $\varepsilon =\Delta $ followed by
saturation at $N_{0}/2$ for $\varepsilon \gg \Delta .$

\bigskip For $h\widetilde{\gamma }>1$\ the minigap at $N_{F\downarrow
}(\varepsilon )$ vanishes and the structure of DoS becomes different, as
illustrated in Fig.4 for the case $h\widetilde{\gamma }=2$. The main
qualitative difference from the previous case is that the spin-down and
total DoS are gapless for $h\widetilde{\gamma }<1$.

\begin{figure}[tbp]
\centerline{\includegraphics[width=85mm]{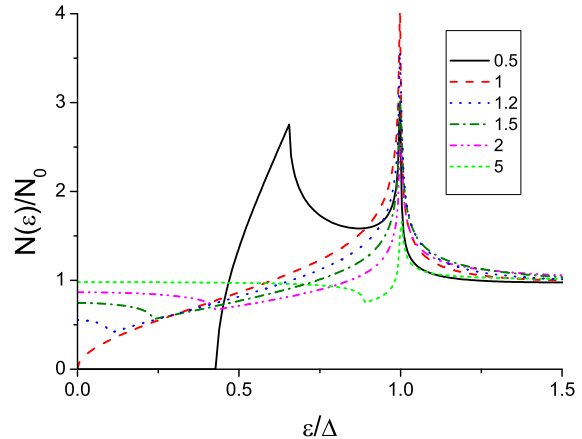}}
\caption{The total DoS in S(FN) for various values of $\widetilde{\protect%
\gamma }h$ as indicated in the figure.}
\label{fig:fig4}
\end{figure}

The total DoS at the F/N interface at $x=0$ (which coincides with the total
DoS in the F/F case), is shown in Fig.5 for various values of $h\widetilde{%
\gamma }.$One can see that the gap is closed at $h\widetilde{\gamma }=1$ and
the broad zero-energy DoS peak is formed with further increase of $h$ until
low-energy states become continuously filled at $h\widetilde{\gamma }\gg 1$.

In conclusion, we have studied theoretically the spatially resolved DoS in
the S(FN) structures and in S(FF) structures with antiparallel magnetization
directions. Analytical solutions were obtained in the case of thin F, N
layers which demonstrate the peculiar features of DoS in this system. We
have illustrated the results numerically and have studied the dependencies
of the minigap and the DoS peak positions on the exchange energy and
parameters of the layers.

This work has been supported in part by Russian Ministry of Education and
Science and RFBR Grant N 04 0217397-a.

\end{document}